\documentclass[iop]{revtex4-1}
\pdfoutput=1 

\usepackage{graphicx} % Include figure files
\usepackage{dcolumn}  % Align table columns on decimal point
\usepackage{bm}       % bold math
\usepackage{amsmath}
\usepackage{amssymb}
\usepackage{hyperref}
\usepackage{epstopdf}

\usepackage{graphicx}

\usepackage[utf8]{inputenc}
\usepackage[english]{babel}
\usepackage{enumerate}

\usepackage{bm}

\begin{document}

\title{Effective equations for reaction coordinates in polymer transport}
\author{Marco Baldovin}
\affiliation{Department of Physics, Universit\`a ``Sapienza'', Roma  
Piazzale A. Moro 5, I-00185 (Italy).}
\author{Fabio Cecconi}
\affiliation{CNR-Istituto dei Sistemi Complessi, Via Taurini 19, I-00185 Roma, 
Italy}
\author{Angelo Vulpiani}
\affiliation{Department of Physics, Universit\`a ``Sapienza'', Roma 
Piazzale A. Moro 5, I-00185 (Italy) and Centro Interdisciplinare ``B. Segre'', 
Accademia dei Lincei, Via della Lungara 10, I-00165 Roma (Italy)}

\date{\today}

\begin{abstract}
In the framework of the problem of finding proper reaction coordinates (RCs)
for complex systems and their effective evolution equations,
we consider the case study of a polymer chain in an external
double-well potential, experiencing thermally activated dynamics.
Langevin effective equations describing the macroscopic dynamics of the system
can be inferred from data by using a data-driven approach, once a suitable set of RCs is chosen.

We show that, in this case, the validity of such choice depends on the stiffness of the polymer's bonds:
if they are sufficiently rigid, we can employ a reduced 
description based only on the coordinate of the center of mass; 
whereas, if the stiffness reduces, the one-variable dynamics is no more Markovian
and (at least) a second reaction coordinate has to be taken into account
to achieve a realistic dynamical description in terms of memoryless Langevin equations.

\end{abstract}

\maketitle %\maketitle must follow title, authors, abstract

%%%%%%%%%%%%%%%%%%%%%%%%%%%%%%%%%%%%%%%%%%%%%%%%%%%%%%%%%%%%%%%%%%%%%%%%%%%%
\section {Introduction
\label{sec:Intro}}
%%%%%%%%%%%%%%%%%%%%%%%%%%%%%%%%%%%%%%%%%%%%%%%%%%%%%%%%%%%%%%%%%%%%%%%%%%%%

The study of many interesting phenomena 
often faces severe difficulties due to the presence of a large amount of  
degrees of freedom and of very different time scales.
As important examples, we can mention protein dynamics and climate physics: 
the time scale of vibrations of covalent bonds is $O(10^{-12}) s$,
while the protein folding time may range from milliseconds to even hours 
for the largest and most complex polypeptides~\cite{fersht99,Finkel,Ghelis2012}.
In the case of climate dynamics the characteristic times 
may range from days (for the atmosphere) to $O(10^4)$ years (for the deep
ocean and ice shields)~\cite{majda06}. In such classes of systems,
numerical simulations are certainly a very  powerful 
and useful tool to investigate the dynamics, but the enormous amount of information 
contained in each single trajectory can be considered somehow 
redundant  if one is interested in a description of the 
processes occurring on a given range of time and spatial scales.
Therefore, the opportunity to use  computational methods
cannot be seen as a ``panacea'' able to explain everything,
because the presence of high-dimensional 
phase-space strongly limits the possibilities to identify and 
extract a simple representation of the relevant processes.

The proper approach in multi-scale systems is the
introduction of suitable 
 effective equations describing  the {\it slow dynamics} in terms of ``slow observables'', 
generally referred to as ``reaction coordinates'' (RC).
RCs are a proper class of observables which are able to characterize, in a reduced way, 
the progress of a reaction in terms of a sequence of chemical events 
(or states)~\cite{schutte}. This methodology is rather useful  both at  practical level 
and from a conceptual point of view:
 effective equations are able to catch certain
general features and to reveal dominant behaviours which could remain
hidden in the fully detailed description~\cite{optimal_RC,schutte}. 
Let us also note that the use of the RCs, which compress (project) the 
multi-dimensional dynamics on a strongly reduced phase-space, 
produces a drastic loss of information; but this loss is compensated 
by an immediate and compact picture of the possible regimes or 
collective behaviors taking place during the system evolution. 
\\
The problem of finding effective equations for multi-scale phenomena has a 
long history in Science, in particular in Mathematics and Physics: as prototype
examples we can mention the averaging method in mechanics~\cite{arnold}
and the Langevin equations for colloidal particles~\cite{castiglione08}.
\\
In few lucky cases, the effective equations can be derived from first 
principles. A remarkable example is the approach dating back to Smoluchowski to 
obtain, using kinetic theory, the Langevin equation for a heavy particle in a 
dilute gas of light particles~\cite{zollette}. Another important attempt was 
suggested by several authors in the 
60's~\cite{rubin60,turner60,mazur-braun64,ford-kac-mazur65} and later by 
Zwanzig~\cite{zwanzig73}, which amounts to rigorously deriving Langevin 
equations for an heavy particle interacting with a chain of light harmonic 
oscillators.

In the study of the dynamical behavior of complex systems with a multi-scale 
structure, the first (and perhaps most difficult) step in the derivation of the 
effective equations, either from first principles or from data analysis, is the identification of suitable RCs. 
This task is far from being trivial and remains conceptually challenging.
We can remind the caveat by Onsager and Machlup in their seminal work on 
fluctuations and irreversible processes~\cite{onsager}: {\it  ``How do you know 
you have taken enough variables, for it to be Markovian?''}
\iffalse
In a similar way, 
Ma~\cite{ma85} noted that {\it ``The hidden worry of thermodynamics is: We do 
not know how many coordinates or forces are necessary to completely specify an 
equilibrium state.''}
\fi

\iffalse
Mori-Zwanzig (MZ) projection, which provides a
systematic method to find an effective stochastic equation for the considered
variable, is one of the most used approaches~\cite{zwanzig61}. In that method the complexity
of the problem is embodied into certain memory kernels, which however can be difficult
to manipulate and to implement numerically.
\fi

There are several systematic methods to partially answer 
the caveat by Onsager and Machlup. The most widely used is \textit{principal
component analysis} (PCA)~\cite{jolliffe16}, which searches for independent linear combinations of
available observables with maximal variance. \textit{Dynamic mode decomposition} (DMD)~\cite{tu14},
\textit{variational approach of conformation dynamics} (VAC)~\cite{noe2013}, \textit{time-lagged
independent component analysis} (TICA)~\cite{molgedey94} are some of the many, related, techniques
used to project the evolution of the coordinates describing the full system
into a smaller set of relevant RCs~\cite{klus18}. This is usually done by considering linear 
combinations of the original variables and exploiting the methods of linear algebra.
In recent years, neural networks and deep learning techniques
have been applied to enhance such algorithms;
specifically, they can select combination of nonlinear functions (from libraries of possible
candidates) to encode original data into the reduced RC space~\cite{wehmeyer16,brunton16}.

 For sure artificial intelligence methods~\cite{russell_AI} are
 useful tools in this perspective. However they 
should not be viewed as automatic or unsupervised 
protocols~\cite{mezard2018,hosni2018,buchanan2019,entropy}:
we cannot disregard the physical intuition and the (partial)
knowledge of the studied systems for selecting the correct RCs
and avoiding ``bad'' choices, which could neglect or hide
relevant phenomenologies occurring in the dynamics.

\iffalse
At a first glance, the recent developments in the field of machine learning 
could raise great optimism about the possibility to build effective equations 
for a large class of systems. Among the others, principal component analysis 
(PCA)~\cite{jolliffe16} and dynamical autoencoders~\cite{wehmeyer16} have shown a 
wide range of application.
\fi

Even in the lucky case in which the proper RCs are already identified, and
we know that the relevant features of the system can be modeled by Markovian
evolution equations for these RCs only, the problem of finding the form of such
equations can be non-trivial. Some methods attempt to extrapolate from data
the most suitable drift and diffusivity 
terms~\cite{peinke, baldolang,entropy} for memoryless 
generalized Langevin Equations. This strategy has been successfully used 
to describe the slow dynamics in several contexts, such as 
turbulence~\cite{renner01, peinke19}, granular media~\cite{baldogran} and 
polymer physics~\cite{hummer2005position, micheletti08}.

In this work, we revisit the Kramers' problem for a polymer in a double 
well~\cite{poly_su_Barr,poly_dwell} by using the evolution equations of 
proper RCs characterizing the slow dynamics of the chain.
We model the polymer as a one-dimensional harmonic chain
(the so-called Rouse chain).

The most natural RCs for our polymer are its center of mass $Q$ and its 
end-to-end distance $L$. We try to derive their evolution equations 
with a data-driven approach: starting from long time series of data,
we apply a well-established procedure to infer the right functional
forms of the Langevin terms appearing in the dynamics. We then analyze
the validity of such description by comparing the behavior of the inferred
model to that of the original system.

The conceptual issues related to the usage of such data-based protocol are discussed.
We shall see that the effective Langevin equations for such RCs depend 
crucially on the stiffness of the polymer bonds: if the polymer is sufficiently rigid, 
only the evolution of $Q$ can be taken into account, while in the regime 
of low stiffness, we need to enlarge the phase space by considering also 
another coordinate ($L$ in our case) to achieve a satisfactory picture of 
the jump dynamics over the barrier. In other words, this second coordinate allows 
the Markov property of the Langevin equations to be preserved.

The outline of the paper is as follows: in Sec.~\ref{sec:system} we describe the 
model; Sec.~\ref{sec:doublewell} reports the 
results about the reconstruction of the Langevin equations obtained by the 
numerical extrapolation procedure, in the case of one RC ($Q$) and in that of two 
RCs (both $Q$ and $L$); finally Sec.~\ref{sec:conclusions} contains our 
conclusions and remarks.

%%%%%%%%%%%%%%%%%%%%%%%%%%%%%%%%%%%%%%%%%%%%%%%%%%%%%%%%%%%%%%%%%%%%%%%%%%%%%5
\section {Model and simple remarks
\label{sec:system}}
%%%%%%%%%%%%%%%%%%%%%%%%%%%%%%%%%%%%%%%%%%%%%%%%%%%%%%%%%%%%%%%%%%%%%%%%%%%%%5
Let us consider the problem of a polymer crossing the barrier of a double-well energy profile,
which is related to the transport of biomolecules accross nano-scale pores~\cite{bezrukov94,kasianowicz96,meller2001voltage,tapio06,ammenti,sebastian06}. 
In many practical situations channels are so narrow that the transport 
dynamics of biopolymers and ions occurs on a single axis, thus, as a matter of fact, it can be 
considered one-dimensional~\cite{cressiot,lubensky1999driven,ansalone,giacometti18}. In this crude 
approximation, the polymer is composed by a chain of $N$ beads (point 
particles), interacting via nearest-neighbors forces and subjected to a 
thermal noise at temperature $T$. 

%--------------------------FIG.1---------------------------------------------
\begin{figure}
\includegraphics[width=.49\linewidth]{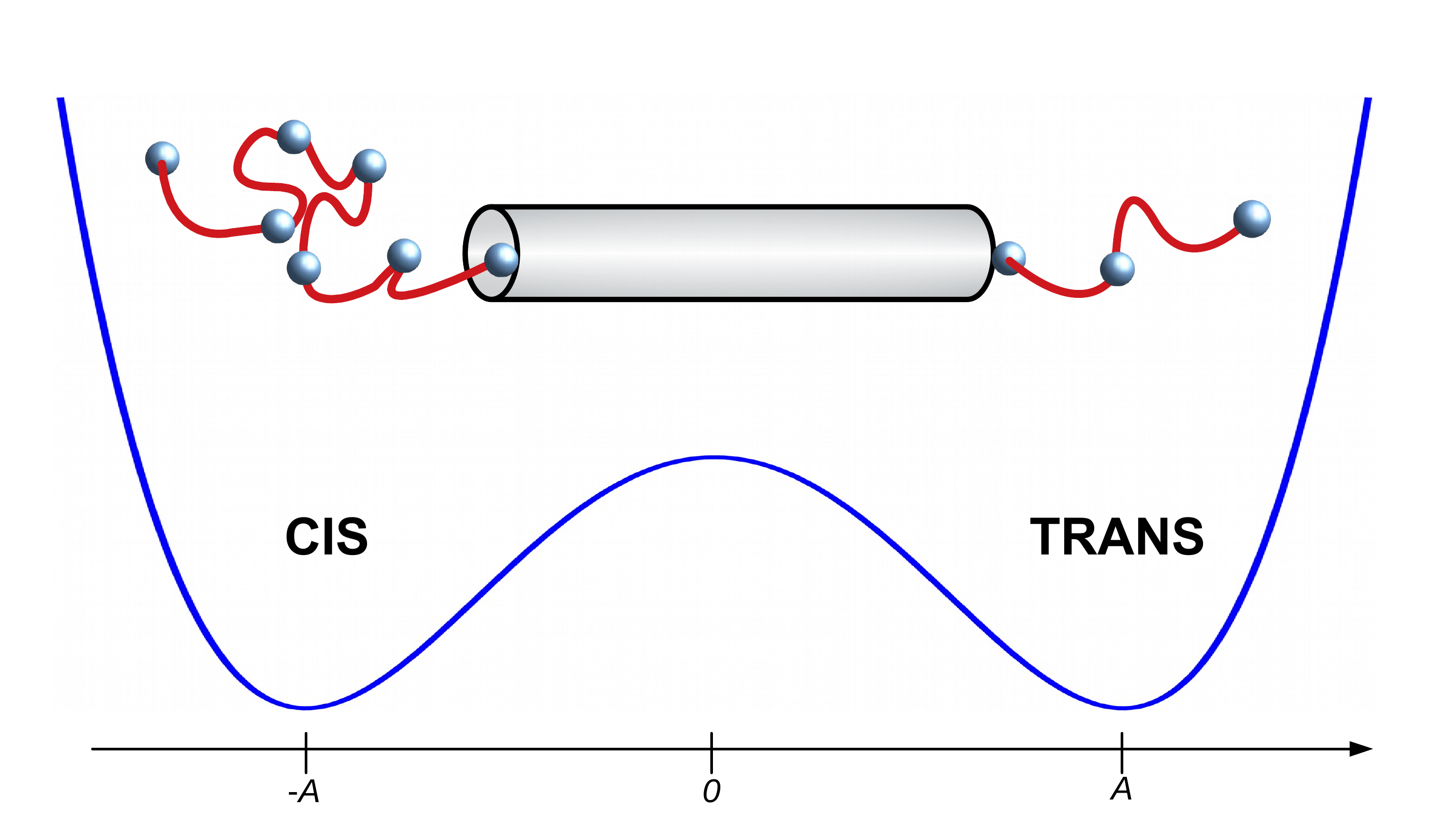}
\caption{\label{fig:nanopore}
 Cartoon of the translocation process of a polymer from CIS to TRANS side of a narrow nanopore.
 The double-well potential is a caricature of the two-state free energy landscape associated to the translocation.} 
\end{figure}
%-----------------------------------------------------------------------------------

The nanopore is portrayed as a region of the 
translocation axis where the polymer feels the effect of an energy barrier, 
which acts independently on each particle and separates the left-side and right-side 
of the pore~\cite{MuthuCapture,ammenti,Polson_Barr}, see Fig.~\ref{fig:nanopore}. 
As customary, in this kind of phenomenology we can assume the evolution
of each monomer to be accessible on time-scales long enough to neglect the
effect of inertia. Accordingly, the monomers are governed by the overdamped
Langevin dynamics:
\begin{eqnarray}
\gamma \dot{x}_1 &=& -V'(x_1) + U'(x_{2} - x_{1}) + \xi_1 \nonumber \\	
\gamma \dot{x}_i &=& -V'(x_i) + U'(x_{i+1} - x_{i}) - U'(x_{i} - x_{i-1}) + 
	\xi_i \nonumber \\
\gamma \dot{x}_N &=& -V'(x_N) - U'(x_{N} - x_{N-1}) + \xi_N
\label{eq:motion}
\end{eqnarray}
with $i=2,\ldots,N-1$, where $x_j$ is the position of the $j$-th bead.
$V$ represents here the external potential, due to 
the nanopore action on the chain. $U$ is the nearest-neighbor 
interparticle potential that is chosen to be 
an even and convex function of $x_{i} - x_{i-1}-\sigma$, 
where $\sigma$ is the equilibrium distance between consecutive particles.
Each $\xi_i$ is a Gaussian noise, with average 
$\langle \xi_i(t)\rangle =0$ and correlation 
$\langle \xi_i(0)\xi_j(t) \rangle = 2 \gamma T \delta_{ij} \delta(t)$, where
$\gamma$ is a dimensional constant, that we will put equal to 1 in the following.

Let us notice, however, that our interest is not in the behavior of this particular
model per se: we are indeed concerned with the general problem of finding proper
RCs and their effective evolution equations from data. In this respect, model~\eqref{eq:motion}
should be meant as a ``generator'' of time series. The analysis is not restricted
to this specific system, and the same conceptual scheme could be applied to more
complex models as well.

As discussed in the Introduction, in order to study the collective dynamics
of the polymer we need to identify
proper reaction coordinates that describe the state of the system, and then
we have to infer effective equations for their evolution.

A natural choice seems to be the center of mass $Q$ of the polymer, which 
roughly indicates the spatial position of the chain:
\begin{equation}
\label{eq:com}
Q=\frac{1}{N}\sum_{i=1}^N x_i\,.
\end{equation}
Its dynamical equation is obtained by just summing up Eq's.~\eqref{eq:motion} 
for all the particles and dividing by $N$,
\begin{equation}
\dot{Q}=-\frac{1}{N}\sum_{i=1}^N V'(x_i) + \sqrt{\frac{2T}{N}}\;\eta_Q
\label{eq:Qdot}
\end{equation}
where the reciprocal elimination of internal forces has been taken into 
account, 
as well as the mutual independence of the noises $\{\xi_i\}$ that combine
into a delta-correlated Gaussian noise with zero mean and such that 
$\langle \eta_Q (0)\eta_Q(t) \rangle = \delta(t)$.

By posing $x_i= Q + u_i$, Eq.~\eqref{eq:Qdot} can be recast as
\begin{equation}
\dot{Q} =-\dfrac{1}{N} \sum_i V'(Q+u_i) + \sqrt{\dfrac{2T}{N}}\;\eta_Q\,.
\label{eq:Qdot1}
\end{equation}
The above equation is formally exact, but it is not very useful in this form, since it depends on
all the $u_i$ terms. 

The roughest approximation that can be done to achieve a closed form for Eq.~\eqref{eq:Qdot1} is to assume that the force term due to the external potential can be written as a (possibly complicated) function of $Q$ only. If this is the case, a one-variable, memoryless model of the kind
\begin{equation}\tag{M1}
\label{eq:M1}
\dfrac{dQ}{dt} = F(Q) +\sqrt{2D}\, \eta_Q\,,
\end{equation}
should catch the relevant features of the macroscopic evolution of the system, where $D=T/N$.
Assuming that the above approximation holds, the specific form of $F(Q)$ can be inferred from data, as will be discussed in the following. 

Let us notice that Eq.~\eqref{eq:M1} describes a Markovian stochastic process for the variable $Q$, and it is expected to give a reasonable approximation of the real dynamics only when the knowledge of $Q$ suffices to determine the macroscopic state of the system. For instance, Eq.~\eqref{eq:M1} gives a good approximation
of the real dynamics in the limit of high-rigidity chain, as it will discussed in the following for a particular case.

In general, however, the above one-variable model will not be valid, meaning that it will not be possible to find any form of $F(Q)$ able to reproduce the dynamical properties of the original system in an accurate way. This is due to the fact that the dynamics of $Q$, in general, is not Markovian: in order to achieve a satisfactory coarse-grained description, one possibility is to modify Eq.~\eqref{eq:M1} by introducing memory-dependent terms, which in some cases can be found analytically by mean of projection methods~\cite{zwanzig61, zwanzig73,grabert80}. To avoid such dependence on memory kernels, which are often difficult to manipulate, the only possibility is to search for (at least) a second RC of the system, such that the vector composed by $Q$ and this new variable obeys a Markovian dynamics.

The additional RC can be individuated through the methods briefly mentioned in the Introduction, or it can be suggested by physical intuition. For instance, one can study a simplified version of the considered problem in order to understand what are the relevant variables.
In our case, if the bond fluctuations are large enough, it is reasonable that the elongation $L=x_N-x_1$ have a role in the macroscopic dynamics. In Appendix~\ref{sec:appendixHCA} we discuss our system under a strong approximation, which allows a simple analytical treatment: in this context the role of $L$ is transparent. 
We can then guess an effective model of the form:
\begin{equation}\tag{M2}
\label{eq:M2}
\begin{cases}	
\dfrac{dQ}{dt} = F_Q(Q,L) +\sqrt{2D_Q}\, \eta_Q\\
\\
\dfrac{dL}{dt} = F_L(Q,L) +\sqrt{2D_L}\, \eta_L\,.
\end{cases}
\end{equation}
Again, assuming that the dynamics of $(Q,L)$ is fairly described by a Markov process, the best choices for $F_Q$, $F_L$, $D_Q$, $D_L$ can be found with data-driven approaches as the one used in this paper. However, one has then to verify that the chosen RCs are actually ``valid'' macroscopic variables, i.e. that the coarse-grained dynamics~\eqref{eq:M2} reproduces the macroscopic features of the original system.

In the remaining part of this paper, we will try to implement this program in a specific case. In particular we will show that, as expected, the stiffness of the polymer plays an important role in the choice of the right set of RCs.

\section{Polymer in a double well
\label{sec:doublewell}}
%%%%%%%%%%%%%%%%%%%%%%%%%%%%%%%%%%%%%%%%%%%%%%%%%%%%%%%%%%%%%%%%%%%%%%%
We consider now the case in which the external potential $V(x)$ in Eq.~\eqref{eq:motion} is a double-well. This 
simple model allows us to study some properties of thermally activated barrier 
crossing as, for instance, the dependence of the jump rate $r$ on the physical 
parameters.

The general problem of activated dynamics has been extensively studied since the seminal works
by Kramers, and the reaction-rate theory provides many analytic methods to
compute jump times in different contexts (see Ref.~\cite{hanggi90} and reference therein).
Important results have been derived also for polymeric chains~\cite{poly_su_Barr,poly_dwell, sebastian06}.
We want to stress that our aim here is not to improve such results: we are interested in
analyzing the ability of a data-driven approach to reconstruct an activated dynamics.
In particular, we will focus on the role of 
the chain stiffness on the activated dynamics and, more importantly, 
on its relevance for the description in terms of the RCs.

The external potential reads, in this case,
\begin{equation}
V(x)=\frac{B^2}{4}(x^2-A^2)^2
	\label{eq:biwell}
\end{equation}
where $A$ and $B$ are suitable constants. The typical dynamics of the center of mass, $Q(t)$,
is therefore characterized by jumps over the barrier separating 
the two minima of the potential (two-state model).  For the interaction potential we choose the form 
$U(r)= K(r - \sigma)^2/2$. 
\\
\iffalse
The description of such transition dynamics strongly depends 
on the bond rigidity $K$. We can expect that in the limit of high
$K$ (semi-rigid rod), the dynamics is fairly reproduced by model (M1). 
As soon as the internal interactions become comparable to the one induced by 
the external potential, a single-variable description fails
and one is forced to consider at least the two-dimensional effective 
dynamics (M2).
\fi

In the following, we will always consider the limit $N\sigma \simeq A$, 
i.e. the case in which the equilibrium length of the polymer is 
comparable to the half distance between the well minima: it is reasonable to 
expect that in these conditions the value of the bond rigidity affects the 
qualitative behavior of the chain in a relevant way.

Our aim is to show that for high values of $K$ the model described by Eq.~\eqref{eq:M1}
suffices to reproduce the quantitative macroscopic behaviour of the system; whereas,
as soon as $K$ becomes comparable to $B^2A^2/\sigma$, the evolution of $Q$ is no more
Markovian and any attempt to describe it through model~\eqref{eq:M1} is doomed to fail.
However, if the phase-space is expanded by including a suitable additional RC, it is
still possible that the evolution of the new RCs vector turns out to be Markovian, so that
a dynamical description based on Eq.~\eqref{eq:M2} can be accurate enough.

The validity of such scenario can be tested by using the data-driven 
approach mentioned in the Introduction and detailed in Appendix~\ref{sec:appendixLE}. 
We first perform numerical simulations of the whole system by using a Stochastic 
Runge-Kutta algorithm~\cite{honeycutt92} and measuring the relevant
RCs of the system at every time step. As a first attempt, from long time-series of
such data we build an effective stochastic equation for $Q$ only,
in the form of Eq.~\eqref{eq:M1}; then we apply the extrapolation procedure to
the dynamics of the two-dimensional vector $(Q,L)$, obtaining an M2-like model.
The ``goodness'' of M1 and M2 is tested
by measuring the Kramers' transition times of the reconstructed 
models and comparing the corresponding jump rates to the original ones.

%%%%%%%%%%%%%%%%%%%%%%%%%%%%%%%%%%%%%%%%%%%%%%%%%%%%%%%%%%%%%%%%%%%%%%%%
\subsection{1-variable model}
%%%%%%%%%%%%%%%%%%%%%%%%%%%%%%%%%%%%%%%%%%%%%%%%%%%%%%%%%%%%%%%%%%%%%%%%

Before applying the mentioned extrapolation method to infer numerically
the functional form of the terms appearing in Eq.~\eqref{eq:M1}, let us derive 
analytically an effective equation for $Q$ for the 
high-stiffness limit, $K\gg B^2A^2/\sigma$. In this case we can assume that
the position of two consecutive beads is fixed and equal to $\sigma$. Due to the 
simple form of the external potential $V(x)$, the drift term in 
Eq.~\eqref{eq:Qdot1} can be exactly computed in this case:
\begin{equation}
\begin{aligned}
 -\frac{1}{N} \sum_{i=1}^N V'(Q+u_i)&= -\frac{B^2}{N} \sum_{i=1}^N \left[ (Q+u_i)^3 -A^2 (Q+u_i)\right]\\
 &= -\frac{B^2}{N} \sum_{i=1}^N \left[ 3Qu_i^2+Q^3-A^2Q\right]\,.
\end{aligned}
\end{equation}
where we have used the fact that, due to the rigidity of the polymer, $\sum_i u_i^3=\sum_i u_i=0$.

Now we substitute the explicit expression for the relative positions of the polymer beads, $u_i=(2i-N-1)\sigma/2$, and after straightforward algebra we get
\begin{equation}
\begin{aligned}
 -\frac{1}{N}\sum_{i=1}^N V'(x_i)&=-B^2 Q\left(Q^2-A^2+\frac{\sigma^2}{4}(N+1)(N-1)\right)\\
 &=-B^2 Q\left(Q^2-A_{\text{eff}}^2\right)
\end{aligned}
\end{equation}
with 
\begin{equation}
 A_{\text{eff}}=A\sqrt{1-\frac{L^2(N+1)}{4A^2(N-1)}}\,,
\end{equation}
where we have used the definition of the polymer length for the rigid case, $L=(N-1)\sigma$.
Let us notice that the above drift corresponds to an effective potential
\begin{equation}
 V_{\text{eff}}(Q)=\frac{B^2}{4}(Q^2-A_{\text{eff}}^2)^2\,,
 \label{eq:poteff}
\end{equation}
i.e. a ``rescaled'' version of the original external potential~\eqref{eq:biwell}.
\iffalse
Let us also remark that, in the limit
$N\gg1$, Eq.~\eqref{eq:closed} (see~\ref{sec:appendixHCA}) is exactly recovered.
\fi
\\
We can now use the theory of escape times~\cite{gardiner} to estimate the jump rate $r$
for the effective potential~\eqref{eq:poteff}. The average waiting time between two consecutive jumps can be computed through the formula~\cite{gardiner}
\begin{equation}
\label{eq:tau_rigid}
\begin{aligned}
 \tau=&\frac{1}{D} \int_{-A}^{A}dy\, e^{V(y)/D} \int_{-\infty}^{y}dz\, e^{-V(z)/D}\\
 =&\frac{1}{D} \int_{-A}^{A}dy\, e^{\frac{B^2}{4D}\left[y^4-2A_{\text{eff}}^2y^2\right]}\int_{-\infty}^{y}dz\, e^{-\frac{B^2}{4D}\left[z^4-2A_{\text{eff}}^2z^2 \right]}
\end{aligned}
\end{equation}
where $D=T/N$ is the diffusivity associated to $Q$. The jump rate $r$ is then found as
\begin{equation}
 r=\frac{1}{\tau}\,.
\end{equation}
The above equations, which are only valid in the rigid-rod limit,
will be a useful touchstone to evaluate the level of accuracy of
the model inferred numerically. 

%------------------------- FIG.2 ------------------------------------------
\begin{figure}
\includegraphics[width=.99\linewidth]{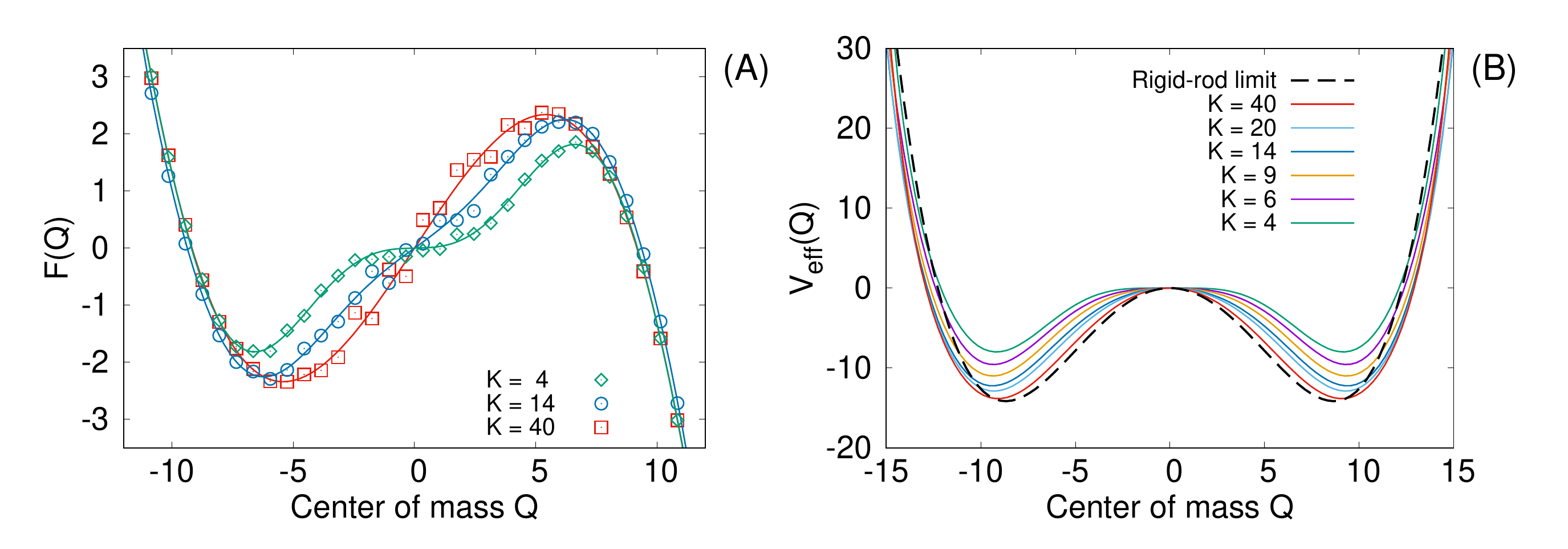}
\caption{\label{fig:potential}
(A): drift term $F(Q)$ of model~\eqref{eq:M1} as reconstructed from data 
(points) and fitted with a 9-th degree odd polynomial (solid lines), for three 
different values of $K$. (B): Effective potential obtained by integration of 
$F(Q)$. Parameters for the 
simulations of the complete system: $A=10$, $B=0.1$, $T=30$, $\sigma=1$, $N=10$, using 
a time-step $dt=10^{-5}$. Simulations on model~\eqref{eq:M1} have been run 
with a time-step $dt=10^{-4}$.} \end{figure}
%-----------------------------------------------------------------------

%---------------------FIG.3--------------------------------------------------
\begin{figure}
\includegraphics[width=.99\linewidth]{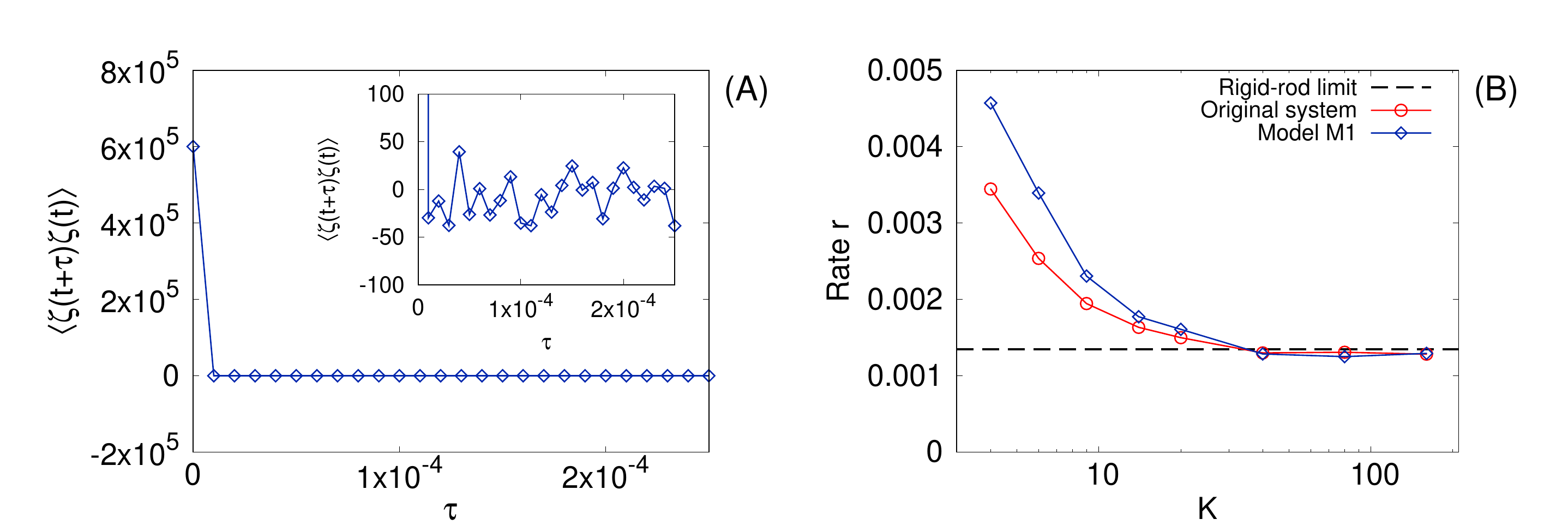}
\caption{\label{fig:markov}
 Checking the validity of model~\eqref{eq:M1}. (A) Autocorrelation $\langle \zeta(t)\zeta(t+\tau) \rangle$ of the ``noise'' term~\eqref{eq:zeta}
vs. $\tau$. The inset is a zoom. It is apparent the fast decay with $\tau$ that can be 
considered effectively a delta-correlation, since the time-step of the
integration algorithm generating the dynamics is $dt=10^{-5}$. Here, $K=14$.
(B): Jump rates as functions of $K$ in the original system (red circles) and in simulations 
of the reconstructed M1 (blue diamonds). The rigid-rod 
approximation (dashed line) is reported as reference. Parameters as in Fig.~\ref{fig:potential}.} 
\end{figure}
%-----------------------------------------------------------------------------------

We apply now the extrapolation procedure mentioned in the Introduction and 
detailed in Appendix~\ref{sec:appendixLE} to infer the right functional forms for 
the terms of the Langevin equation~\eqref{eq:M1}, assuming that the dynamics of $Q$
is Markovian and a 1-variable description in the form of model~\eqref{eq:M1} holds.
We find that $D$ (not shown here) is always almost constant and equal to $T/N$,
as it would be expected if the process was Markovian, while the drift shows a 
more complex shape (Fig.~\ref{fig:potential}A); we fit the data by a 9-th degree, 
odd polynomial, then we integrate the resulting function in order to get an 
effective potential, which is reported in Fig.~\ref{fig:potential}B for several 
values of $K$. In the large-$K$ limit, as expected, we recover a quartic 
effective potential: terms of higher order become relevant when the bond stiffness is 
low, and their effect is to flatten the potential barrier between the two wells.

As mentioned above, the validity of Eq.~\eqref{eq:M1} relies on the assumption that
the evolution of $Q$ is Markovian, which has to be checked.
First, one can define and measure the following quantity:
\begin{equation}
\label{eq:zeta}
 \zeta(t)=\dot{Q}(t)-F(Q(t))\,,
\end{equation}
which represents the ``noise'' of Eq.~\eqref{eq:M1}, if the dynamics of $Q$ is Markovian. We can compare the autocorrelation time of $\zeta(t)$ and verify that it is much shorter than any characteristic time-scale of the dynamics of $Q$. In our case, $\zeta(t)$ always decorrelates on the scale of the time-step of the integration algorithm, $dt$ (see Fig.~\ref{fig:markov}A).

This first check assures that there is a clear time-scale separation between the dynamics of the center of mass and its ``noise''. However, this does not imply that the original dynamics of $Q$ has to be Markovian: in order to check that, we also have to verify the consistency with the original dynamics. If the one-variable description is able to catch the relevant features of the whole system, we can conclude, \textit{a posteriori}, that the evolution of $Q$ was Markovian also in the complete dynamics; if not, a different description has to be taken into account.

Fig.~\ref{fig:markov}B shows, for several values of 
the rigidity, the jump rates measured in the original dynamics and those 
observed in the reconstructed model, using a standard stochastic integration 
algorithm (the one discussed in~\cite{mannella-palleschi89}, up to order 
$dt^{3/2}$). In the high-$K$ limit the simple rigid-rod 
approximation~\eqref{eq:tau_rigid} holds, there is no dependence on $K$ and the 
agreement between the jump rates of M1 and of system~\eqref{eq:motion} is excellent. As the polymer 
becomes softer, even if a significant improvement on Eq.~\eqref{eq:tau_rigid} 
can be observed, the relative error between M1 and the true dynamics exceeds 
30\%: this is a clear hint that a 1-variable description, even if inferred 
directly from data, cannot reproduce all the relevant features of the dynamics.
This is due to the fact that our implicit assumption on the Markovianity of the
process is wrong.

%=======================================================================
\subsection{2-variables model}
%=======================================================================

The failure of model~\eqref{eq:M1} for small values of $K$,
revealed by the discrepancies between the reconstructed and the 
original jump rate, suggests to go beyond a 
single variable description in order to achieve satisfactory results.
As discussed in Section~\ref{sec:system}, a reasonable attempt to recover
a Markovian dynamics is to consider the elongation of the polymer, $L$,
as a second $RC$ for our model, and we postulate the validity of an 
evolution equation of the form~\eqref{eq:M2}.
\iffalse
The same also happens under the simplifying
assumption of homogeneous deformations, which makes the problem exactly 
solvable, see Appendix~\ref{sec:appendixHCA}.
\fi

\begin{figure}[h!]
\includegraphics[width=0.5\linewidth]{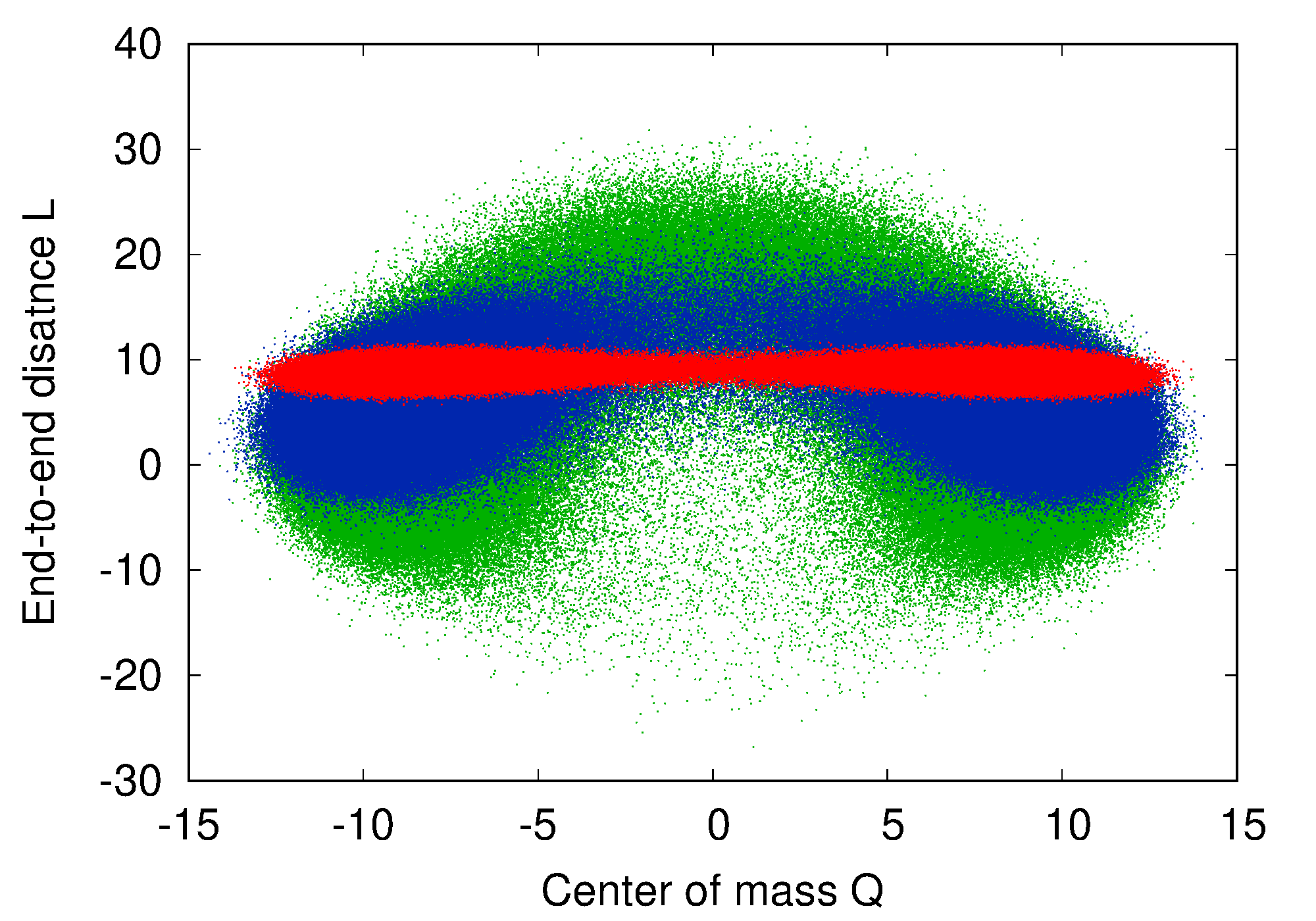}
\caption{\label{fig:scatter}
Scatter plot of the end-to-end distance $L$ vs. $Q$ (total integration time: $10^6$). Green dots: $K=4$; blue dots: $K=20$; red dots: $K=600$. Other parameters as in Fig.~\ref{fig:potential}.}
\end{figure}

\begin{figure}[h!]
\includegraphics[width=.8\linewidth]{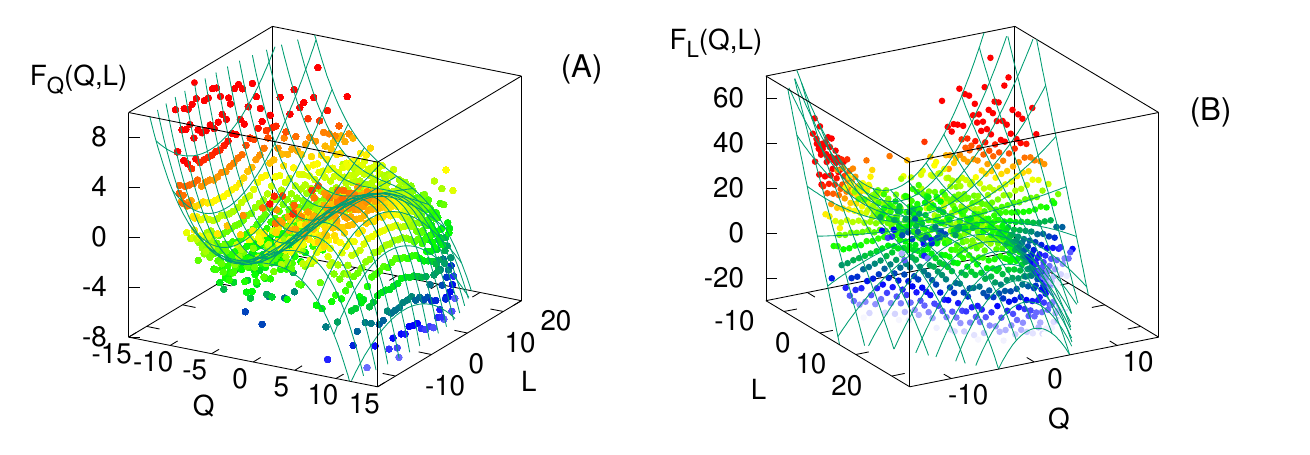}
\caption{\label{fig:fit_3d}
	Reconstructed drift terms of $Q$ and $L$ in the model \eqref{eq:M2}, 
	case $K=4$. Points are extrapolated from data (see Appendix~\ref{sec:appendixLE}); the 
	surface is obtained by fitting the polynomial~\eqref{eq:FQ_FL}. Other parameters are as in Fig.~\ref{fig:potential}.}
\end{figure}

The requirement of a variable accounting for the elongation of the polymer
can be easily understood by looking at Fig.~\ref{fig:scatter} reporting
three scatter plots of the original dynamics in the $(Q,L)$ plane,
for different bond rigidity, where $L = x_N - x_1$.
When $K$ is high, and the system is well approximated by the rigid-rod model,
the region of the phase space explored by the dynamics is a thin strip
around the equilibrium value $L\simeq (N-1) \sigma$. 
As soon as the rigidity condition
is relaxed, and the system is allowed to vary its length in a significant way,
a two-lobe distribution takes place: $L$ tends to be smaller than the rest 
length of the chain when the polymer occupies one of the two minima of the double-well
potential, while it significantly increases during the transition across the barrier.
This particular shape of the scatter plot indicates that 
the typical pathways in the space $(Q,L)$ include a non-negligible 
deformation in $L$, which can be straightforwardly interpreted as follows: when the rigidity $K$ 
is low, the transition across the barriers of the polymer
occurs with a concomitant stretching of the bonds, presumably those that instantaneously lay on top the barrier. 
As a consequence, any Markovian effective description involving only the center of mass 
is completely insufficient to fairly approximate the dynamics of the system. 
%------------------------------- FIG.4 ------------------------------------
%-------------------------------------------------------------------------
Following again the strategy discussed in Appendix~\ref{sec:appendixLE}, 
we provide numerical values for $F_Q$, $D_Q$, $F_L$ and $D_L$ in the $(Q,L)$ 
space,
which have to be fitted using suitable functional forms. Due to the symmetries
of the system, $F_Q(Q,L)$ has to be odd with 
respect to the variable $Q$, while $F_L(Q,L)$ should be even. 
Fig.~\ref{fig:fit_3d} shows the results obtained by fitting the following 
polynomial form:
\begin{equation}
\begin{aligned}
F_Q(Q,L)& = Q\left[c_{10}^{(Q)}   + c_{12}^{(Q)}L^2   + c_{13}^{(Q)}L^3\right] \\
	& + Q^3\left[c_{30}^{(Q)} + c_{32}^{(Q)}L^2 + c_{33}^{(Q)}L^3\right]\\
F_L(Q,L)& = c_{00}^{(L)}    + c_{01}^{(L)}L      + c_{03}^{(L)}L^3
	  + c_{21}^{(L)}Q^2L\,.
\end{aligned}
\label{eq:FQ_FL}
\end{equation}

%------------------------------ FIG.5 --------------------------------------
%---------------------------------------------------------------------------

The agreement between the actual data and the proposed functional form is good enough
to hope that the guessed model catches the most relevant features of the dynamics.
The diffusivity terms $D_Q$ and $D_L$ are again fitted by constant functions.
Once model~\eqref{eq:M2} is determined, 
we can check the reliability of its stochastic evolution
by a direct comparison with the original dynamics.

%---------------------------- FIG.6 --------------------------------------
\begin{figure}[h!]
\includegraphics[width=.5\linewidth]{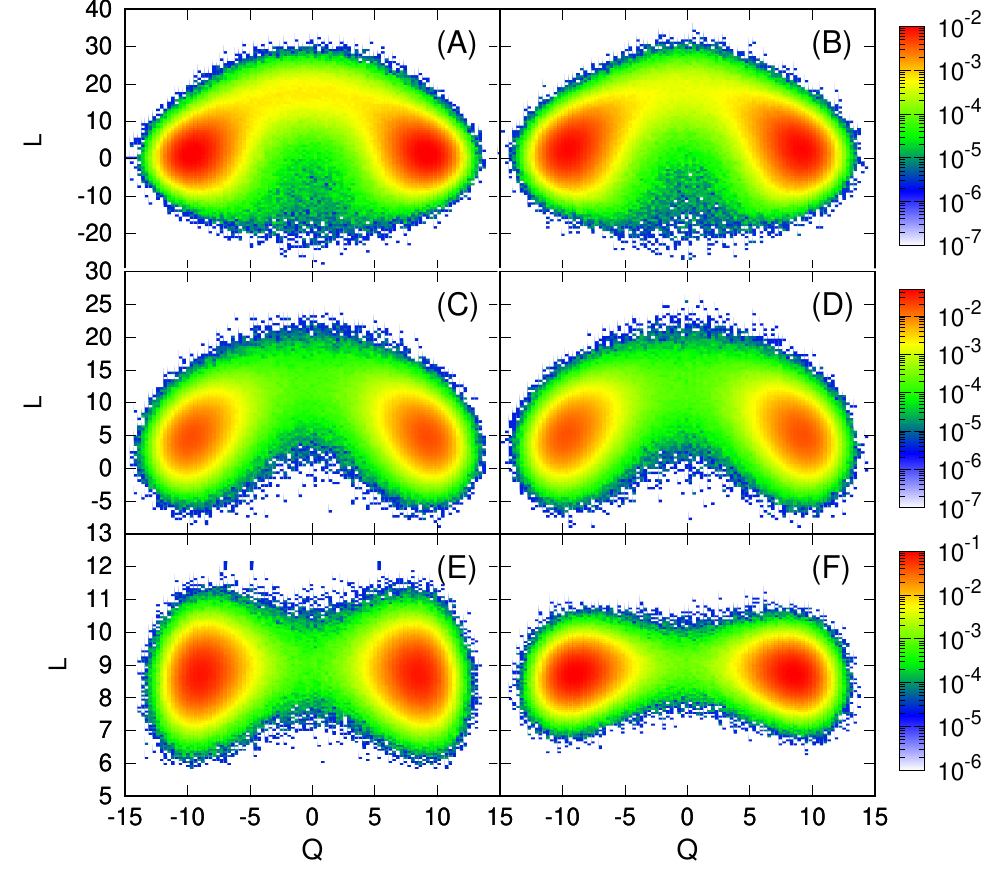}
\caption{\label{fig:pdfs}
Pdfs in the 2-variable phase space of the original system, first column, 
and in the reconstructed 2-variables model~\eqref{eq:M2}, second column. 
Stiffness: $K=4$ (top), $K=20$ (center), $K=600$ (bottom).  Other parameters
are the same as Fig.~\ref{fig:potential}.}
\end{figure}
%-------------------------------------------------------------------------

A first, important benchmark is given by the ability of the model to reproduce 
the static properties of the system, namely the joint probability distribution 
in the $(Q,L)$ space. This test is reported in Fig.~\ref{fig:pdfs} for different 
values of $K$, showing a reasonable qualitative agreement even in the non-trivial
case of low bond stiffness: in particular, the stretching occurring when the polymer
crosses the barrier is clearly reproduced.
%----------------------------- FIG.7 -----------------------------------
\begin{figure}[h!]
\includegraphics[width=0.5\linewidth]{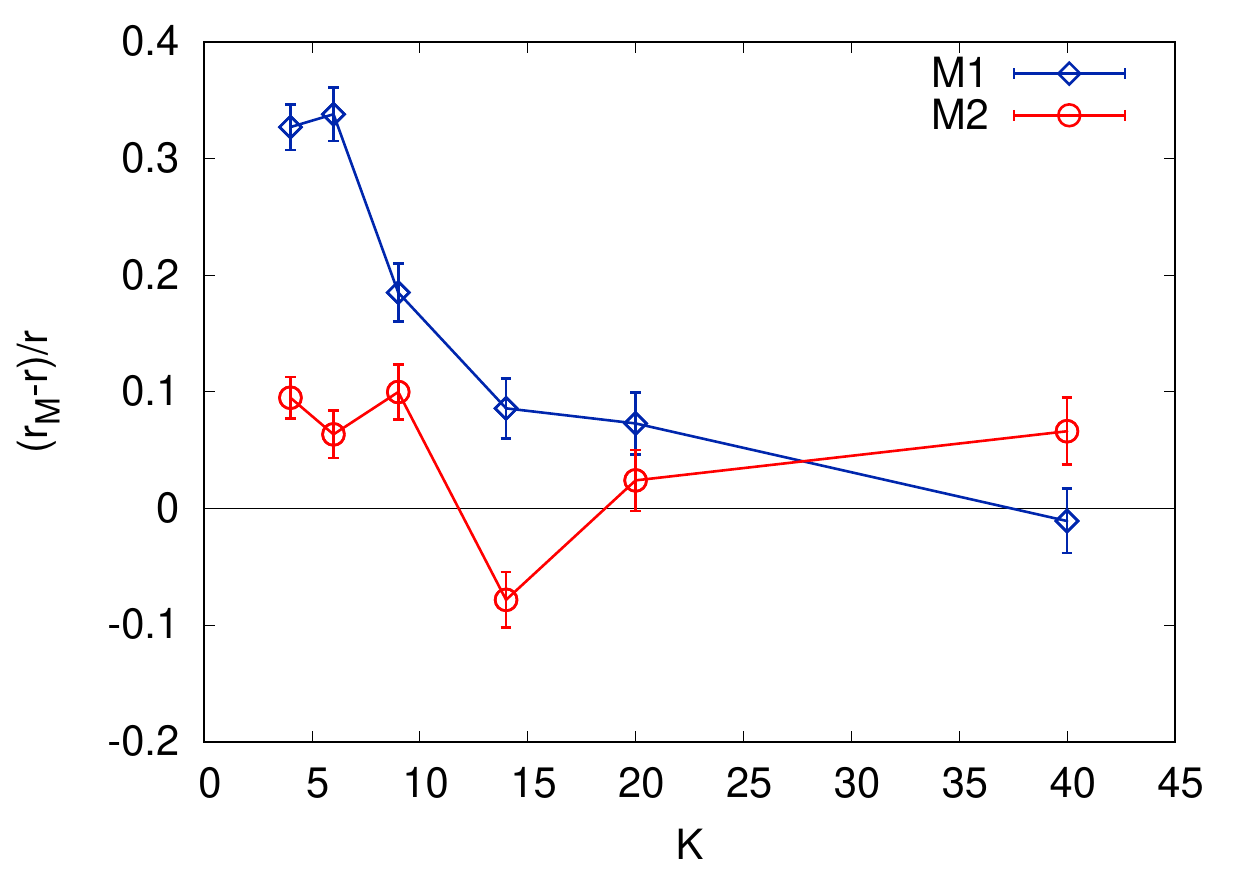}
\caption{\label{fig:rates}
Relative errors of the jump rates $r$ in the two reconstructed models,
varying $K$. For other parameters, see caption of Fig.~\ref{fig:potential}.}
\end{figure}
%-----------------------------------------------------------------------
The improvement of our effective description when also $L$ is taken into 
account can be fully appreciated by looking at dynamical observables
as the jump rate $r$. 
Fig.~\ref{fig:rates} displays the relative errors
between the values of $r$ obtained in the reconstructed models M1, M2 and
in the original dynamics. As already discussed, 
the 1-variable model fails when the polymer is soft, while the accuracy
of M2 does not seem to be affected in this limit.
Let us notice, on the other hand, that for $K\gg A^2B^2/\sigma$ the reconstructed model~\eqref{eq:M2} is less reliable than the 1-variable version: this is probably
a consequence of the larger number of parameters involved, which leads to a lower
degree of precision on their determination with the discussed method.

%=============================================================================
\subsection{Remarks on the structure of the effective equations}
%=============================================================================
The procedure for the reconstruction of a model in the form~\eqref{eq:M2} that 
we used in the previous subsection is based on a ``dynamical'' analysis, which 
builds the coefficients of the guessed stochastic model by looking at the 
time evolution of suitable observables of the original system (see 
Appendix~\ref{sec:appendixLE}). 
One could wonder whether such procedure is really 
needed in order to get a realistic description of the studied process; for 
example, in analogy with many statistical mechanical problems, one may expect
that the following recipe works:
\begin{enumerate}
\item measure the stationary p.d.f. $\rho(Q,L)$ from long time series of data;
\item deduce an effective 2-variable ``potential'', also known as ``potential
of mean force'' in chemical and biophysical contexts:
\begin{equation}
\label{eq:pot_stat}
W_S(Q,L)= -\log[\rho(Q,L)]\,;
\end{equation}
\item define $F_Q\equiv-c\partial_Q W_S$ and $F_L\equiv-c\partial_L 
W_S$, where the constant $c$ has to be determined.
\end{enumerate}

Let us notice that in many practical situations the multidimensional free-energy landscapes obtained by simulations or experiments are assumed to be generated by a system with a gradient (or gradient-like) structure and, using this hypothesis, Langer's formula~\cite{langer69,hanggi90} is applied to derive the transition rates over saddles.

The above procedure is a completely ``static'' analysis,
because it involves only quantities measured in equilibrium conditions.
Leaving apart the problem of finding the right multiplicative constant $c$
and the noise terms, this approach has a 
major issue: it is not sure, \textit{a priori}, whether the dynamics of 
the chosen reaction coordinates can be described by a potential, 
even if the complete system actually can; in other words,
there is in general no reason to expect that the model is a gradient system in the reduced phase space.
\iffalse
This gradient structure, however, is not guarateed \textit{a priori} and it needs to be carefully checked: the case discussed in this section is a clear example in which this assumption fails and one cannot rely on the mentioned approach.
\fi

The knowledge of $W_S$ alone is not sufficient to obtain the drift; additional informations on the dynamics need to be taken into account. A possibility is the theoretical approach discussed in Ref.~\cite{grabert06}, which also involves the transport coefficients. Our numerical method, instead, exploits the dynamical information by computing suitable conditioned moments of the RCs, as discussed in Appendix~\ref{sec:appendixLE}.

%------------------------------ FIG.8 ------------------------------------
\begin{figure}[h!]
\includegraphics[width=.5\linewidth]{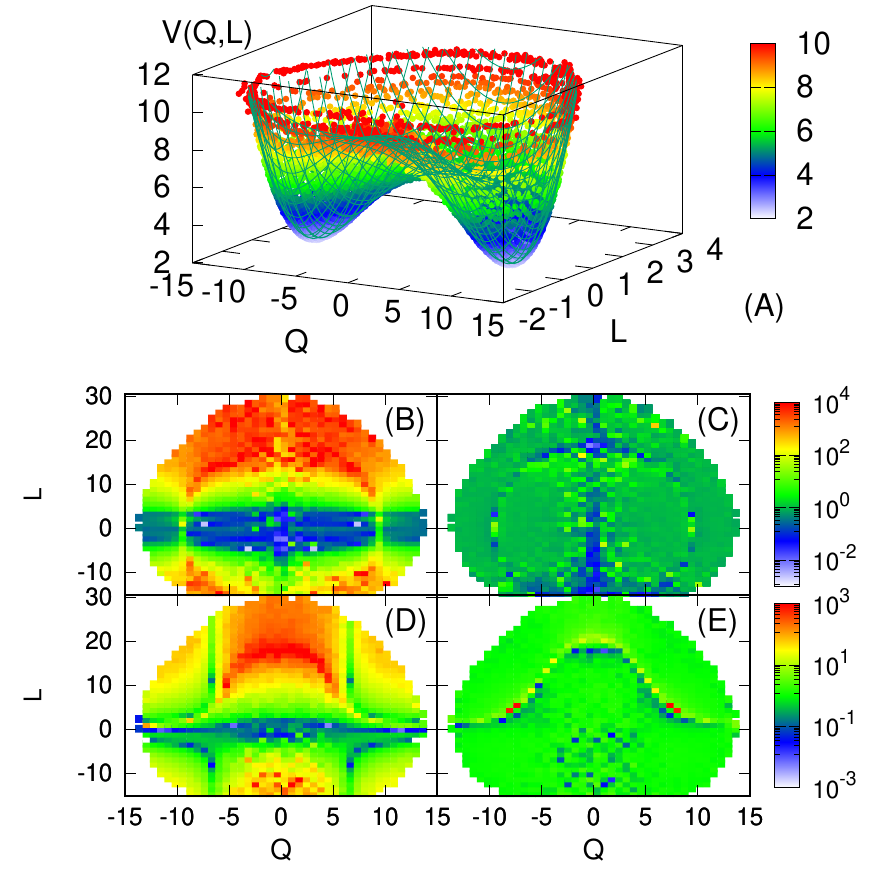}
\caption{\label{fig:stat}
Static analysis. (A): effective potential $W_S(Q,L)$ from the empirical pdf in 
the $(Q,L)$ space. Panels below compare the drift computed with this static approach,
$\nabla W_s$, and the one reconstructed with our dynamical approach, $(F^{(2)}_Q, F^{(2)}_L)$,
to the measured coefficients $F_Q, F_L$. (B): $|\partial_Q W_s/F_Q|$. (C): 
$|F^{(2)}_Q/F_Q|$. (D): $|\partial_L W_s/F_L|$. (E): $|F^{(2)}_L/F_L|$.} 
\end{figure}
%----------------------------------------------------------------------
In order to show that in our case the simplifying assumption of a gradient structure 
for the $(Q,L)$ dynamics is wrong, let us determine $W_S$ using Eq.~\eqref{eq:pot_stat}. 
In Fig.~\ref{fig:stat}A we fit such ``potential'' with a 2-variable polynomial
(4-th order in both $Q$ and $L$), getting a nice superposition.

If the 2-variables system were gradient, $F_Q(Q,L)$ should be equal to $c\partial_Q W_S$,
for some value of $c$; therefore the ratio $\partial_Q W_S/F_Q$
should be constant all over the phase-space. Fig.~\ref{fig:stat}B shows instead that, in our case,
such function strongly depends on $Q$ and $L$. 
Just for comparison, Fig.~\ref{fig:stat}C displays the ratio 
$$
\epsilon =  F^{(2)}_Q/F_Q
$$ 
where $F^{(2)}_Q$ represents here the expression from the fitting 
\eqref{eq:FQ_FL}: not surprisingly, this function
is constant and equal to one almost everywhere, meaning only that we have made
sensible choices for the polynomial functions to use in the fit.
In Fig.~\ref{fig:stat}D,~\ref{fig:stat}E  the same comparison is done for the drift of the 
end-to-end distance $L$.

This simple check clearly shows that the simplified 2-variables system is not gradient,
and therefore a static analysis
is not sufficient to infer reasonable effective equations for its evolution.

%%%%%%%%%%%%%%%%%%%%%%%%%%%%%%%%%%%%%%%%%%%%%%%%%%%%%%%%%%%%%%%%%%%%%%%%%%%%%%
\section{Conclusions}
\label{sec:conclusions}
%%%%%%%%%%%%%%%%%%%%%%%%%%%%%%%%%%%%%%%%%%%%%%%%%%%%%%%%%%%%%%%%%%%%%%%%%%%%%%
We studied the dynamics of a 1-dim polymer whose monomers are subjected to a
double-well external potential. Under certain conditions, the phenomenology is characterized
by thermally activated barrier crossing (classical Kramers' problem). 
We addressed the issue of describing this complex high-dimensional 
dynamics in terms of few suitable observables, the reaction 
coordinates (RCs). 
In our system the center of mass, $Q$, and the end-to-end distance, $L$, are the 
most natural candidates.

These RCs evolve according to effective Langevin equations that are 
generally difficult to be derived via a systematic procedure.
The proper reconstruction of the effective stochastic equations for $Q$ and 
$L$ is achieved via a data-driven numerical method that extrapolates 
the drift and diffusion terms from a long trajectory of the original system.
Let us stress that this method allows us to find nonlinear terms for the
reconstructed Langevin equation; this is an important difference, e.g., from
the standard Mori-Zwanzig approach, where the complexity of the problem
is shifted into the shape of the memory kernels.~\cite{zwanzig61}
The reliability of the reconstructed dynamics is tested on the way it 
fairly reproduces some essential properties of the original dynamics, 
such as a correct estimate of the jump rate over the barrier.

From our study it emerges that the description level in terms of RCs strongly 
depends on the bond stiffness $K$. More precisely: if the bonds are 
rigid enough, we are allowed to consider the evolution of $Q$ only, given by 
model \eqref{eq:M1}, to fully characterize the jump dynamics.
However, when $K$ decreases, the internal motion of the 
chain cannot be neglected and also the dynamics of $L$ has to be considered, 
so that a satisfactory description can only be obtained in the plane $(Q,L)$ 
through the model \eqref{eq:M2}. On a more mathematical perspective, lowering $K$ can be regarded as a loss of Markovianity of the one-variable description. A second coordinate is necessary to recover the Markov property.
\iffalse
The emergence of such qualitative scenario can be also shown analytically,
under the simplifying assumption of HCA.
\fi

Our work shows how subtle is the procedure of reducing the dynamics of a
many-dimensional system to a low-dimensional model, even in the simplest cases 
where the physical intuition leaves little ambiguity to the choice of the RCs.
%Under other physical conditions this choice cannot be 
%satisfactory. 
In fact, the choice of an RC can be correct in certain regimes but not 
sufficient in others. Specifically in our case, we can only know 
a posteriori that it is the stiffness of the polymer to determine 
whether one or two RCs are needed.

\appendix

%%%%%%%%%%%%%%%%%%%%%%%%%%%%%%%%%%%%%%%%%%%%%%%%%%%%%%%%%%%%%%%%%%%%%%%%%%%%
\section{The homogeneous chain approximation}
\label{sec:appendixHCA}
%%%%%%%%%%%%%%%%%%%%%%%%%%%%%%%%%%%%%%%%%%%%%%%%%%%%%%%%%%%%%%%%%%%%%%%%%%%%

In this Appendix we discuss the ``homogeneous chain approximation'' (HCA), which amounts to assuming that the distances
between nearest-neighbor particles, at each time $t$, are all equal:
\begin{equation}
\label{eq:hca_approx}
 x_{i+1}(t)-x_i(t) = x_{j+1}(t)-x_j(t)\,,\quad \forall i,j\,.
\end{equation}

Under such approximation we can derive closed
analytical expressions for both models~\eqref{eq:M1} and~\eqref{eq:M2};
HCA is very unphysical, and it only holds true if the polymer bonds are almost rigid.
However, a simple analysis of this limit can give us some insight on the general case.
Even under such simplifying hypotheses, as soon as $K$ is large enough, model~\eqref{eq:M1}
is not sufficient to describe correctly the dynamics, and a~\eqref{eq:M2}-type model is needed.

First, let us consider the case of high intra-chain forces acting on each monomer.
Specifically, the bonds are much stronger than: i) the external forces due to the action of the potential 
$V(x)$ and ii) those induced by the thermal fluctuations. 
Under conditions i) and ii) our polymer 
reduces to a rigid rod with fixed distances among its elements, i.e.
\begin{equation}
 x_{i+1}(t)-x_i(t)=const \quad \forall i\,.
\end{equation}

Eq.~\eqref{eq:Qdot1} can be written as
\begin{equation}
\label{eq:rho}
\dot{Q} = -\int\!\!du \rho(u) V'(Q+u) +
\sqrt{\dfrac{2T}{N}}\;\eta_Q\;.
\end{equation}
where we have introduced the density,
$\rho(u) = 1/N \sum_{i=1}^N \delta(u-u_i)$, to pass from the sum to an 
integral. Here
\begin{equation}
\rho(u)\simeq\frac{1}{L}\Theta(u^2-L^2/4)\,,
\end{equation}
where we also took the $N\gg1$ limit.
%\begin{equation}
%-\dfrac{1}{N} \sum_i V'(Q+u_i) \simeq \frac{1}{L} \int_{-L/2}^{L/2}\!\!du V'(Q+u)\,
%\label{eq:homog_approx}
%\end{equation}
The total length of the polymer is constant, $L=(N-1)\sigma$.  

In this simple case, we can straightforwardly apply the fundamental theorem
of calculus and get:
\begin{equation}
\dot{Q} = \dfrac{1}{L}[V (Q - L/2) - V(Q + L/2)] +
\sqrt{\frac{2T}{N}}\eta_Q\,.
\label{eq:closed}
\end{equation}
Notice that the above equation is of the form of model~\eqref{eq:M1}.

Let us consider now
the idealized situation in which the inter-particle distances do fluctuate, 
but the approximation~\eqref{eq:closed} still holds because the polymer 
undergoes homogeneous deformations. 
In this limit, the end-to-end distance $L$ defined as
\begin{equation}
\label{eq:defL}
 L=x_N-x_1
\end{equation}
is no longer constant and Eq.\eqref{eq:closed} 
should be complemented by a new equation describing the dynamics of $L$.
Now $L$ is a necessary second RC, so
the phase space is enlarged to the plane $(Q,L)$. 

Under this hypothesis, we can derive a second equation, by 
writing $\dot{L} = \dot{x}_{N}- \dot{x}_{1}$ from Eq.~\eqref{eq:motion}
and then approximating each $x_{i}- x_{i-1} \simeq L/(N-1)$.
The final result for the drift expressions is 
\begin{eqnarray} 
F_Q(Q,L)& = & \dfrac{1}{L} \left[V\left(Q-\dfrac{L}{2}\right) - 
 	 V\left(Q+\dfrac{L}{2}\right)\right] \nonumber \\
	& & \label{eq:QL_evol}\\
F_L(Q,L) & = & -2 U'\left(\frac{L}{N-1}\right) + 
	 V'\left(Q - \frac{L}{2}\right) - V'\left(Q+\frac{L}{2}\right)
	    \nonumber
\end{eqnarray}

The above phenomenological discussions suggest that there are two 
regimes depending on the polymer stiffness:
\begin{itemize}
 \item stiff chain: the polymer dynamics can be characterized by 
	 a single reaction coordinate $Q$, for which a closed evolution 
	 equation can be found; 
 \item soft chain: also a second variable, for instance 
	 the end-to-end distance $L$, is needed to close the evolution 
	 equations.
\end{itemize}

Notice that in the above discussion the introduction of $L$ as a second $RC$
emerges quite naturally. This can be a clue on the relevant variable to choose
also in the more general case where assumption~\eqref{eq:hca_approx} does not hold.

\iffalse
It is important to stress that the necessity of a 
second RC is actually a general condition to achieve a 
correct description of the chain dynamics when the bonds
between the monomers are weak, even under the HCA approximation.
\fi

%%%%%%%%%%%%%%%%%%%%%%%%%%%%%%%%%%%%%%%%%%%%%%%%%%%%%%%%%%%%%%%%%%%%%%%%%%%%
\section{Extrapolating Langevin equations from data}
\label{sec:appendixLE}
%%%%%%%%%%%%%%%%%%%%%%%%%%%%%%%%%%%%%%%%%%%%%%%%%%%%%%%%%%%%%%%%%%%%%%%%%%%%
In this Appendix we briefly recall the basic aspects of the extrapolation procedure that
we use to infer the parameters of effective Langevin equations from long-time series of
data (in this case, produced by numerical simulations). An extensive discussion of the method
can be found in~\cite{peinke, entropy} and reference therein. See also~\cite{baldogran} for a case in which
the study of a multi-dimensional system is considered.

Let us assume that each component of the vector variable $\mathbf{X}$ obeys the following Langevin's equation:
\begin{equation}
 \dot{X}_i=F_i(\mathbf{X})+\sqrt{2D_i(\mathbf{X})}\xi_i\,
\end{equation}
where each $\xi_i$ is a white noise with unitary variance. For the sake of simplicity, here we assume that the coefficients $F_i$ and $D_i$ do not depend on time.
It can be proved~\cite{peinke2} that the following relations hold:
\begin{equation}
\begin{aligned}
 F_i(\mathbf{x})&=\lim_{\Delta t \to 0} \frac{1}{\Delta t}\langle \Delta X_i | \mathbf{X}(0)=\mathbf{x}\rangle\\
 D_i(\mathbf{x})&=\lim_{\Delta t \to 0} \frac{1}{2\Delta t}\langle \left(\Delta X_i-F_i(\mathbf{x})\Delta t\right)^2 | \mathbf{X}(0)=\mathbf{x}\rangle
\end{aligned}
\label{eq:d&d}
\end{equation}
where $\Delta X_i=X_i(\Delta t)-X_i(0)$.
Due to the stationarity of the process, we can compute the ensemble averages~\eqref{eq:d&d} as temporal averages over long-time series of data.

The $\Delta t \to 0$ limit has to be interpreted in a proper physical way: for every real phenomenon, a stochastic description holds only for some not-too-small time scales. It is customary to define a typical time-scale of the considered problem, usually referred to as the ``Markov-Einstein time'' $\tau_{ME}$, such that the Langevin equations hold true only if one considers time scales larger than $\tau_{ME}$. In order to get a reasonable esteem of the above limits, a good strategy is that of plotting the conditioned moments on the r.h.s. of Eq.~\eqref{eq:d&d} as functions of $\Delta t$, then to individuate a sufficiently regular region that allows for a $\Delta t \to 0$ extrapolation. In our case, however, the separation of time-scales is a consequence of the fast decorrelation times of the quantity $\zeta(t)$ defined by Eq.~\eqref{eq:zeta}, which is in turn due to the overdamped nature of the original dynamics we are considering. As a consequence, it is sufficient to take $\Delta t \gg dt$, where $dt$ is the time-step of the integration algorithm. Let us notice once again that such time-scale separation does not imply the Markovianity of the considered process, and the validity of such approximation can be only checked \textit{a posteriori}.

%------------------------------ FIG.9 ------------------------------------
\begin{figure}[h!]
\includegraphics[width=0.8\columnwidth]{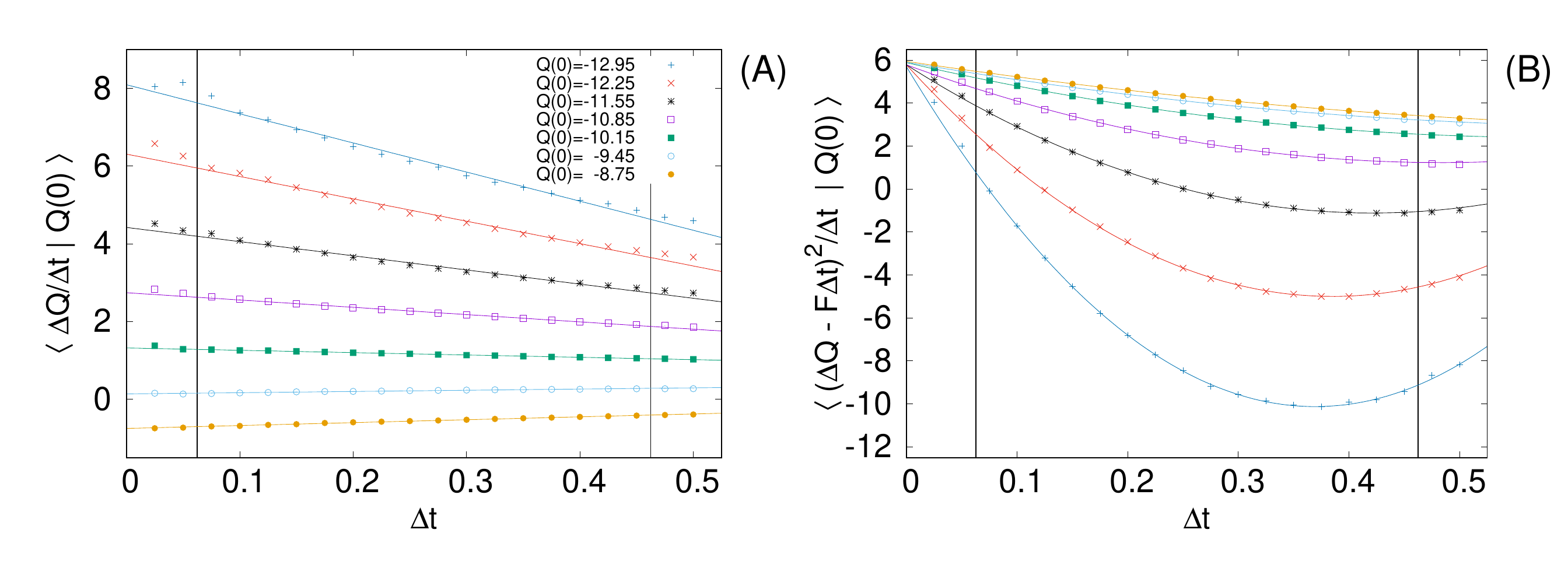}
\caption{\label{fig:extrap}
Extrapolation of the drift (panel A) and diffusivity (panel B) from data. The quantities in the rhs of Eq.~\eqref{eq:d&d} are fitted with a low-order polynomial; only the regions within the vertical bars are considered for the fit. The $\Delta t \to 0$ limit is then taken as the vertical intercept of the fitted functions with the $y$-axis. Here $K=4$ (other parameters as in Fig.~\ref{fig:potential}).
}
\end{figure}
%----------------------------------------------------------------------

In Fig.~\ref{fig:extrap} we show a typical case of reconstruction of the drift and diffusivity terms of model~\eqref{eq:M1}, for several values of the center of mass.

\begin{acknowledgments}
This work is part of MIUR-PRIN2017 \textit{Coarse-grained description for non-equilibrium systems and transport phenomena (CO-NEST)}. 
\end{acknowledgments}

\bibliography{biblio}

\end{document}